\documentclass[a4paper]{JHEP3}
\usepackage{amsmath,amssymb,amscd}

\title{%
\protect\vspace{5mm}
On Existence of Self-Tuning Solutions in
Static Braneworlds without Singularities
}
\author{
Peter Koroteev$^{abc}$ and Maxim Libanov $^{a}$
 \\
$^{a}$~Institute for Nuclear Research of the Russian Academy of
Sciences,\\
60th October Anniversary Prospect 7a, Moscow 117312 Russia\\
$^{b}$~Moscow Institute of Physics and Technology\\
Moscow 141701, Russia\\
$^{c}$~Institute for Theoretical and
Experimental Physics\\
Moscow 117218, Russia\\
E-mail: \email{koroteev@ms2.inr.ac.ru, ml@ms2.inr.ac.ru}
}
\preprint{
ITEP-TH-60/07 \\
ULB-TH/07-34
}
\abstract{
A static self-tuning $SO(3)\times\mathbb{Z}_2$ symmetric and translation
invariant braneworld setup with flat brane is considered. We discuss the
null energy conditions (NEC) for matter on the brane and in the bulk and
prove that for the static regular background with broken Lorentz
invariance the NEC and positiveness of the total energy density on the
brane and NEC in the bulk cannot be satisfied simultaneously. Then we give
some examples and elaborate some special cases. For instance, we provide a macroscopic solution for a background with Lifshitz scaling.
}

\begin{document}
\section{Introduction and Motivation}

Models with large extra dimensions were suggested considerably long time
ago as possible solutions of fundamental problems of cosmology and
particle physics like cosmological constant problem, hierarchy problem or
problem of generations (see e.g. Ref.~\cite{ArkaniHamed:1998rs}; for a
review see Ref.~\cite{Rubakov:2001kp} and for earlier works see
Refs.~\cite{Rubakov:1983bb, Antoniadis:1990ew}). In the recent eight years
braneworld models have attracted very much attention because of works
\cite{Randall:1999ee, Randall:1999vf} by L.~Randall and R.~Sundrum (we
will refer to these models as to RS1 and RS2 models correspondingly). In
these models it has been shown that once a four dimensional flat brane is
embedded into a five dimensional anti de-Sitter space then static
gravitational potential on a brane at small distances is exactly Newtonian
up to  a power correction with a coefficient depending on a warping
factor. Therefore at small energies an observer does not see extra
dimension and does not detect deviations from a four dimensional physics.
The interest to the braneworld models is also warmed up by upcoming high
energy experiments on LHC.

In the RS2 model a bulk is filled by a negative cosmological constant and
no other media are considered. In many further contributions different
braneworld backgrounds were elaborated and different bulk and brane types
of matter were investigated (see for review e.g.~\cite{Brax:2003fv,
Kiritsis:2003mc}). One expects to have a configuration with a ``physical''
matter, that is the one with an energy density being positively defined
and a null energy condition (NEC) being satisfied. A natural question
arises whether this kind of solutions exists. The answer to this question
for a generic (even $\mathbb{Z}_2$ symmetric) braneworld setup with time
dependent metric is, however, a formidable task as it entails rather
complicated Einstein equations to be solved. Thus first one can simplify
the problem by asking the same question for static configurations where
rigorous analysis can be performed. This is the main goal of the present
work. We consider static backgrounds with a broken five-dimensional
Lorentz invariance. Our Universe is supposed to be located on a 4-d brane
representing a slice in a five dimensional spacetime.

The motivation of this paper is twofold. The first one comes from the
so-called no-go theorem discussed in \cite{Cline:2001yt}. The authors
prove that if a static braneworld solution has a singularity at some point
in a bulk then this singularity is necessarily naked (that is not screened
by any horizon from the brane) if NECs both on the brane and in the bulk
are satisfied. The existence of the singularity was a starting point of the
proof. But still the following question remains unclear whether there
exists a self-tuning static setup with nonsingular metric provided that
null energy conditions are satisfied on the brane and in the bulk.

The second motivation comes from \cite{Libanov:2005yf, Libanov:2005nv}
where the authors in a context of a so-called ``trans-Planckian problem''
use '\textit{ad hoc}' Lorentz violating background and do not provide the
reader with energy-momentum tensor which solves field equations. A natural
question arises whether the corresponding matter satisfies NEC or not. In
the current work we answer both questions from the papers
\cite{Cline:2001yt, Libanov:2005yf, Libanov:2005nv} and generically both
are
negative if the static background is considered. As an answer to the first
question we prove a theorem that no non-singular solutions are possible
thereby answering the second question posed in
\cite{Libanov:2005yf, Libanov:2005nv} where the setup is assumed to be
smooth and non-singular\footnote{It is worth to stress here, however,
that in~\cite{Libanov:2005yf, Libanov:2005nv} non-static metric has been
considered, so our present result can be viewed only as an indication that
the setup in~\cite{Libanov:2005yf, Libanov:2005nv} is pathological.}.

The paper is organized as follows. In the Section \ref{sec:setupformul} we
spell out the metric tensor, bulk Einstein equations, impose junction
conditions at the brane location, discuss null energy conditions and
formulate the main statement. The Section \ref{sec:proof} is devoted to
proof of the this statement. Then in the Section \ref{sec:examplesevade}
we discuss various examples illustrating the theorem and specify the
conditions under which the theorem can be evaded. As one of the examples we construct a macroscopic solution for a background with Lifshitz scaling. In the end we make some
concluding remarks.

\section{The Setup and the No--Go Theorem}\label{sec:setupformul}

First we  specify the metric we are going to employ in the current paper.
The following coordinates are used: $t,\,\textbf{x}=(x^1,x^2,x^3)$ are
four-coordinates parallel to the brane and $z$ is the bulk coordinate with
$z$-axis being orthogonal to the brane. We impose $SO(3)\times
T^3\times\mathbb{Z}_2$ symmetry which acts on spatial coordinates
$(x^1,x^2,x^3)$ by rotations and translations, and $\mathbb{Z}_2$
reflection of $z$ axis as $z\mapsto -z$ (the latter is imposed for
simplicity). Moreover we shall consider only flat brane. One can show that
under these symmetries the general static metric can be written as
follows\footnote{Indeed, $SO(3)$ rotations fix $\textbf{x}$ dependence of
  the functions $a$ and $b$ as follows: $a(z,x)=a(z,|\textbf{x}|)$ and
$b(z,x)=b(z,|\textbf{x}|)$. The requirement that the spacetime in question
is translation invariant along $(x^1,x^2,x^3)$ rules out $x$ dependence
completely.}
\begin{equation}
\label{eq:generalansatz}
ds^2 = e^{-2a(z)}dt^2 -
e^{-2b(z)}d\textbf{x}^2-dz^2
\end{equation}
The brane is located at $z=0$ and the functions $a(z)$ and $b(z)$ are
assumed to be $z$ even. For each fixed value of $z$ the Universe
represents a slice warped by factors $e^{-2a(z)}$ and $e^{-2b(z)}$. If the
functions $a$ and $b$ are different then Lorentz invariance in the bulk is
violated. We shall also impose the condition that at the location of the
brane at $z=0$ Lorentz invariance is reproduced and by choosing an
appropriate normalization one can put $a(0)=b(0)=0$ thereby setting the
metric to the Minkowskian one at this point.

\paragraph{Einstein Equations and Junction Conditions.}

The bulk Einstein equations relating Einstein tensor $G_{AB}$ and energy
momentum tensor read\footnote{For simplicity we put the coefficient in
front of the energy-momentum tensor to be equal to 1 since it does not
play any role in our further discussions.}
\begin{equation}
\label{eq:EinsteineqnsGen}
G^A_B = T^A_B+\delta ^{A}_{B}\Lambda ,
\end{equation}
where $T^A_B$ is the bulk energy-momentum tensor and $\Lambda$ is the
cosmological constant which is assumed to be negative ($\Lambda <0$).
The non zero Einstein tensor components for the metric
\eqref{eq:generalansatz} are the following
\begin{align}
\label{eq:EinstenTensorGen}
G^0_0&=3b''-6b'^2\notag\\
G^1_1&=2b''+a''-3b'^2-a'^2-2a'b'\\
G^5_5&=-3b'^2-3b'a'\notag
\end{align}

We introduce Minkowski brane at $z=0$ with the following five-dimensional
energy-momentum tensor
\begin{equation}
\label{eq:energymomentumbrane}
T^A_{B,\,b}=\mbox{diag}(\rho_b+\sigma,\,\,-p_b+\sigma,\,\,-p_b+\sigma,
\,\,-p_b+\sigma,\,\,0)\,\delta(z)
\end{equation}
where $p_{b}=\omega_{b}\rho_{b}$ is the brane equation of state and
$\sigma$ is the brane tension. Israel junction condition
\cite{Israel:1966} applied to our case reads
%
\[
[K^A_B]\Big\vert_{z=0}=\int\limits_{-\infty}^{+\infty}\Big(T^A_{B,\,b}-
\frac{1}{3}T^C_{C,\,b}h^A_B\Big)dz,
\]
%
where we put the extrinsic curvature junction at the brane location in the
l.h.s and $h_{AB}$ is the induced metric on the brane. Applying the above
condition to our metric \eqref{eq:generalansatz} we obtain
\begin{align}
\label{eq:IsraelOur}
-2a'(0)&=p_{b}+\frac{2}{3}\rho_{b}-\frac{\sigma}{3},\notag\\
-2b'(0)&=-\frac{\rho_{b}}{3}-\frac{\sigma}{3}.
\end{align}
From these equations brane matter quantities can be extracted.

\paragraph{The Null Energy Condition and the No-Go Theorem.}

A NEC can be expressed in the covariant way via
energy momentum tensor as follows
\begin{equation}
\label{eq:NECGen}
T_{AB}\xi^A \xi^B\geq 0,\quad g_{AB}\xi^A \xi^B=0
\end{equation}
for every null vector $\xi^A$. Physically a NEC implies that in adiabatic
media a speed of sound does not exceed a speed of light.

Let us first sketch a couple of examples. Given a theory of an
usual scalar field the NEC is satisfied automatically. For the perfect
fluid with pressure $p$, energy density $\rho$ and state equation $p=w\rho$
it reads $w>-1$ and $w=-1$ for a vacuum.

The statement we prove in the current paper approves the result obtained in
\cite{Cline:2001yt}. The authors state that it is impossible to shield the
singularity from the brane by a horizon, unless the NEC is violated in the
bulk or on the brane. Here we are going to prove the following statement:

\noindent
\textit{Let the following conditions hold
\begin{enumerate}
\item bulk NEC\,\, \, \, $T_{AB}\xi^A \xi^B\geq 0,\quad g_{AB}\xi^A
\xi^B=0$
\item brane NEC\,\,\,\,  $T_{b,\,\mu\nu}\xi^{\mu} \xi^{\nu}\geq 0,\quad
g_{b,\,\mu\nu}\xi^\mu \xi^\nu=0$
\item positiveness of brane energy density\,\,\,\, $\rho_b+\sigma \geq 0$
\item 3d brane flatness\,\,  $k=0$
\item bulk Lorentz invariance violation\,\, $a(z)\neq b(z)$.
\end{enumerate}
Then and only then a generic background of type \eqref{eq:generalansatz}
without bulk singularities\footnote{By singularity a physical one is
assumed, i.e. the case when the energy--momentum tensor has poles.} does
not exist.}

Note that the conditions 1,2 and 3 in the above theorem are physical but
the conditions 4 and 5 might be relaxed in principle. In the later case the
theorem does not hold and we shall provide the reader with corresponding
examples further.

\section{The Proof}\label{sec:proof}
Here we adduce the rigorous proof of the statement from previous Section
which can be treated as a mathematical theorem. We note that it holds only
for generic $a$ and $b$. The special case $a=b$ corresponding to the
Lorentz invariant setup enables us to evade it under certain choice of the
function and will be considered later.

\paragraph{The NEC inequalities.}

The bulk NEC reads $\mathcal{T}=T_{AB}\xi^A \xi^B\geq 0$ or equivalently
\[
T^0_0 g_{00}(\xi^0)^2+\sum\limits_{i=1}^3 T^i_i g_{ii}(\xi^
i)^2+T^5_5 g_{55}(\xi^5)^2\geq0
\]
where the vector $\xi^A$ lives on the 5D light cone $g_{00} (\xi^0)^2+\sum
g_{ii} (\xi^i)^2+g_{55} (\xi^5)^2=0$. Thus in order to provide the above
inequality one can find an absolute minimum of the bilinear
form $\mathcal{T}(\xi )$
\begin{equation}
\mathcal{T}=g_{00}T^0_0(\xi^0)^2- |g_{11}| T^1_1\sum\limits_{i=1}^3
(\xi^ i)^2-T^5_5(\xi^5)^2
\label{eq:Tform}
\end{equation}
on the light cone
\begin{equation}
g_{00} (\xi^0)^2-
|g_{11}|\sum(\xi^i)^2=(\xi^5)^2
\label{eq:constraint}
\end{equation}
and require that $\mathcal{T}_{\mathrm{min}}\geq 0$.

Expressing $\xi^0$ from \eqref{eq:constraint} and substituting it into
\eqref{eq:Tform} one has
\[
\mathcal{T}=|g_{11}|\Big( T^0_0- T^1_1\Big)\sum\limits_{i=1}^3(\xi^
i)^2+\Big( T^0_0- T^5_5\Big)(\xi^5)^2
\]
The form $\mathcal{T}$ is nonnegatively defined iff the inequalities
$T^0_0- T^1_1\geq 0$ and $T^0_0- T^5_5\geq 0$ are satisfied. Using the
Einstein equations (\ref{eq:EinsteineqnsGen}) and Einstein tensor
(\ref{eq:EinstenTensorGen}) one can express these inequalities in terms of
the metric coefficients
\begin{align}
b''-a''-3b'^2+a'^2+2a'b'&\geq0
\label{eq:necbulk1}\\
b''-b'^2+b'a'&\geq0
\label{eq:necbulk2}
\end{align}
These constraints should be completed by the brane NEC which can be
extracted from Israel junction condition \eqref{eq:IsraelOur}. Subtraction
of the second equation from the first one in \eqref{eq:IsraelOur} yields
%
\[
2(b'(0)-a'(0))=p_b+\rho_b
\]
%
We also consider $\rho_b+\sigma \geq 0$. Thus the brane NEC and
positiveness of brane energy density imply that at $z=0$ we have
\begin{equation}
b'(0)-a'(0)\geq 0,\quad b'(0)\geq 0
\label{eq:braneNEC}
\end{equation}
which means that these inequalities are also satisfied at least at some
vicinity of $z=0$.

Reshuffling \eqref{eq:necbulk1} and \eqref{eq:necbulk2} one has
\begin{align}
b''-a''-3(a'-b')^2-4a'(b'-a')&\geq0
\label{eq:f1}\\
b'' - (b'-a')^2-a'(b'-a') &\geq0
\label{eq:f2}
\end{align}
Thus the bulk and the brane NECs can be reformulated as
\eqref{eq:braneNEC}--\eqref{eq:f2}.

\paragraph{The proof.}

Now we shall consider all possible cases assuming smoothness of the
functions $a(z)$ and $b(z)$ and figure out whether it is possible or not
to satisfy \eqref{eq:braneNEC}, \eqref{eq:f1} and \eqref{eq:f2}
simultaneously. Let us first assume that in \eqref{eq:braneNEC}
$b'(0)>a'(0)$. Then there exists some positive $z_c$ and $0< z< z_c$ such
that the inequality \eqref{eq:braneNEC} holds in this vicinity. In this
region the inequality \eqref{eq:f1} can be replaced by the following
equation
\begin{equation}
\label{eq:f1eq}
b''-a''-3(a'-b')^2-4a'(b'-a')=\phi(b'-a')
\end{equation}
where $\phi=\frac{T_0^0 - T^1_1}{b'-a'}$ is a nonnegative function for
$0<z<z_c$ and has to be finite (as we will see $b'(z)\neq a'(z)$) for
regular solutions. (The fact that $\phi\geq 0$ is crucial in the proof
whereas its finiteness will not be used further). The equation
(\ref{eq:f1eq}) can be solved in terms of $b'$ generically as
follows\footnote{We are going to utilize this trick in several places in
this proof. The main idea is to show that under imposed conditions a
denominator inevitably becomes zero at some point in the bulk thereby
turning the whole expression to infinity. It shows that no background
without bulk singularities is allowed under imposed bulk and brane NECs.}
\begin{equation}
\label{eq:bpmapint}
b'(z)=a'(z)-\frac{\exp{\Big(4a(z)+4\int\limits_0^z
\phi(y)\,dy\Big)}} {3\int\limits_0^z
dy\,\exp{\Big(4a(y)+4\int\limits_0^y\phi(t)\,dt\Big)}-C}
\end{equation}
and boundary conditions fix the constant of integration
$1/C=b'(0)-a'(0)\geq 0$. Special case $b'(0)=a'(0)$ we shall consider
later. Since $b'-a'\geq 0$ for $0<z<z_c$ then the function $b-a$ increases
at this interval. Let us first assume that $z_c$ is finite and $b'-a'\leq
0$ for $z>z_c$. The function $b-a$ stops increasing and starts decreasing
at $z_c$ which means that the fraction in the r.h.s of \eqref{eq:bpmapint}
becomes zero at $z_c$. This is potentially possible in the two following
cases:
\begin{itemize}
 \item [1)] the numerator vanishes;
 \item [2)] the denominator turns to infinity.
 \end{itemize}

The case $1)$ cannot be realized since in order to make the exponent in
the numerator to be equal to zero (as the integral
of $\phi$ is always nonnegative) $a$ has to have a pole $a\to-\infty$ at
$z_c$. It means that the metric becomes singular and no smooth solution is
possible. The case $2)$ can be excluded as well because  of the following
considerations. The integral $\int_0^{z_c}
dy\,\exp{\Big(4a(y)+4\int\limits_0^y\phi\, dt\Big)}$ has to diverge at
$z_c$. It means there exists some point $z_{\ast}$ from $0<z_{\ast}<z_c$
such that $\int_0^{z_{\ast}} dy\,\exp{\Big(4a(y)+4\int\limits_0^y\phi \,
dt\Big)}=C$. This entails a pole for $b'-a'$ according to
\eqref{eq:bpmapint}. Hence this case is also not allowed.

Thus we showed that $b-a$ has to increase at the whole half-line. Let us
consider the inequality \eqref{eq:necbulk2}. Analogously with
\eqref{eq:necbulk1} it can be reformulated as an equation with some
nonnegative function $\chi$ involved
\begin{equation}
\label{eq:bpint}
b'(z)=\frac{\exp{\Big(-a(z)+\int\limits_0^z
\chi(y)\,dy\Big)}}{-\int\limits_0^z dy\, \exp{\Big(-a(y)+\int\limits_0^y
\chi(t)\,dt\Big)}+\tilde C}
\end{equation}
Here $\tilde C=1/b'(0)>0$. (The case $b'(0)=0$ is possible as well.
Because of \eqref{eq:braneNEC} it follows that $a'(0)=0$ also.
This case will be elaborated further.) Taking into account
\eqref{eq:bpmapint} and \eqref{eq:bpint} and applying the considerations
we used after \eqref{eq:bpmapint} one can see that in order to keep the
functions $b'-a'$ and $b'$ finite in the bulk the following integrals
%
\[
\int\limits_0^\infty e^{4a(z)} dz <\infty,\quad\text{and}\quad
\int\limits_0^\infty e^{-a(z)} dz <\infty
\]
%
have to converge which cannot be provided simultaneously.

Now we treat the case with $b'(0)=a'(0)$ (it also includes the case
$b'(0)=0$). One can have $b'(z)=a'(z)$ for all positive $z$ or $b'(z)\neq
a'(z)$ in some domain. The first opportunity will be considered separately
in the Section \ref{sec:examplesevade} and corresponds to the Lorentz
invariant case. As for the second one it appears that only $b'(z)\geq
a'(z)$ case can be realized. Indeed let us perform Taylor expansion of
both functions
\begin{equation}
a'(z)=a_0 + a_1 z + O(z^2),\quad b'(z)=a_0+b_1 z + O(z^2).
\label{Eq/Pg10/1:NEC}
\end{equation}
The constant $a_0$ appears in both expansions and provides the condition
$b'(0)=a'(0)$. Then up to the first order in $z$ the inequalities
\eqref{eq:f1} and \eqref{eq:f2} become
%
\[
b_1-a_1-4a_0(b_1-a_1)z+O(z^2)\geq 0,\quad
b_1-a_0(b_1-a_1)z+O(z^2)\geq 0
\]
%
We assume that $a_0\neq 0,\,a_1\neq 0,\,b_1\neq 0$. It follows that
$b_1\geq 0$ and $b_1\geq a_1$ in order to provide \eqref{eq:f1} and
\eqref{eq:f2} at small $z$. Hence there should exists some point $z_n >0$
where $b'(z_n)> a'(z_n)$ as the functions $b'$ and $a'$ are assumed to be
different.\footnote{If all derivatives in the function $b$ (or $a$) up to
n'th order vanish one should start the expansion in (\ref{Eq/Pg10/1:NEC})
with $b_n+b_{n+1}z^{n+1}+\dots$ and proceed analogously.} Starting from
this point we can solve the same Cauchy problem as \eqref{eq:f1eq} but on
the interval $(z_n,+\infty)$ with initial conditions
$b'(z_n)-a'(z_n)=\text{const}>0$. All considerations can be repeated and
one can show that $b'-a'$ will diverge at some point in the bulk.

So far we have elaborated all possible cases except the one with
$a(z)=b(z)$ corresponding to the Lorentz invariant background. If this is
not the case then the statement from the last passage of the Section
\ref{sec:setupformul} is proved. But it appears that for $a=b$ there
exists a class of solutions which respect NEC. We shall consider it
separately in the next Section.

\paragraph{Are the found singularities physical?}

Now we show that the singularities we have found above are indeed physical,
i.e. the energy momentum tensor has to become infinite at these points.
The verification should be done as it happens sometimes that a
singularity is just coordinate one (like horizon point in
Schwarzschild-type metrics).

Let us consider $T^5_5$ which by corresponding Einstein equation is equal
to $G^5_5$ from \eqref{eq:EinstenTensorGen}
%
\[
T^5_5 = -3b'(b'+a')
\]
%
If $b'$ has a pole and $a'$ is finite or vice versa then $T^5_5$ has a
pole as well. As we have shown the functions $a'$ and $b'$ cannot have
singularities at the same point otherwise one of the expressions
\eqref{eq:bpmapint} or \eqref{eq:bpint} becomes infinite. The remaining
case is when $a'=b'$ identically and it can be also ruled out as $T^5_5$
diverges. Thus we have shown that all singularities are essentially
physical and correspond to poles in the energy-momentum tensor.

\section{Examples, Special Cases and How to Evade the Theorem}
\label{sec:examplesevade}
In this Section we discuss various deviations from the statement we have
just proved. The first one refers to the Lorentz invariant case when the
metric can be rewritten via one warp factor
\begin{equation}
\label{eq:metricLI}
ds^2=e^{-2b(z)}(dt^2-d\textbf{x}^2)-dz^2.
\end{equation}
We recall that the case with $b(z)=kz$ for $k=\sqrt{-\Lambda/6}$
corresponds to the well known RS models \cite{Randall:1999ee,
Randall:1999vf}. The null energy condition for this case is satisfied as
$\Lambda$ term does not affect the inequality~\eqref{eq:NECGen}. We are
going to show that one can satisfy NEC as well with some media in the bulk
apart from cosmological constant for the background \eqref{eq:metricLI}
under particular choice of the function $b$.

\paragraph{Lorentz invariant case.}

For the Lorentz invariant case bulk and brane NECs simplify dramatically.
The inequality \eqref{eq:necbulk1} becomes trivial and \eqref{eq:necbulk2}
(which comes from the condition $T_{0}^{0}-T_{5}^{5}\geq 0$) reads simply
as $b''\geq 0$. Thus if we do not require anything else here then the NEC
in the bulk can be satisfied with ease as well as all the other four
conditions in the theorem. Due to physical reasons let us consider a bulk
matter with nonnegative energy density $T^0_0\geq 0$ (note that in the
case with broken Lorentz invariance this condition is redundant). Using
\eqref{eq:EinsteineqnsGen} and \eqref{eq:EinstenTensorGen} one has
%
\[
3b''-6b'^2-\Lambda=T^0_0\geq 0
\]
%
We are going to apply the same trick here and rewrite the inequality via
the following equation with some function $\phi$
%
\[
b''\Big(1-\frac{\Lambda}{3b''}\Big)-2b'^2-\phi b'=0
\]
%
where we have identified $3\phi b'=T^0_0$ and assumed that there exists a
region where $b''$ does not vanishes. The assumption is reasonable as due
to \eqref{eq:f2} which for the Lorentz invariant case turns merely into
$b''>0$ at least in some vicinity of the origin or otherwise $b''=0$ for
all $z$ (RS2 solution).\footnote{The same considerations as in the
previous passage may be applied to show that if $b''(0)=0$ then it will
either be identically zero or it will be positive at least in small
vicinity of the origin.} We have
%
\[
b'=\frac{\exp{\Big(\int\limits_0^z
dy\,\frac{\phi(y)}{1-\Lambda/3b''}\Big)}}{-2\int\limits_0^z
dy\,\frac{\exp{\Big(\int\limits_0^y
dt\,\frac{\phi(t)}{1-\Lambda/3b''}\Big)}}{1-\Lambda/3b''}+C'}
\]
%
One can see that if the integral in the denominator of the above
expression converges as $z\to\infty$ then the appropriate initial
condition $C'$ in a way to make the denominator nonzero thereby providing
finiteness of $b'$ in the bulk can be chosen.

For instance the following function can be considered
%
\[
b''=\left\{
\begin{aligned}
(z-z_0)^2,\quad &0\leq z<z_0,\\
0, \quad &z\geq z_0.\\
\end{aligned}
\right.
\]
%
For $z\geq z_0$ the solution corresponds to the RS2 case and $T^A_B$
vanishes but for $0\leq z<z_0$ one has different solution with nonzero
bulk matter energy-momentum tensor satisfying the NEC.

\paragraph{Example with Lifshitz scaling.}

Here we shall discuss one parametric deformation of the RS2 model with
broken Lorentz invariance. The following energy-momentum tensor for the
bulk matter is used
\begin{align}
\label{eq:EMTperfectfluid}
T^0_0&=((1+\omega)u_0u^0-\omega)\rho\notag\\
T^1_1&=((1+w)u_1u^1-w)\rho\\
T^5_5&=((1+\omega)u_5u^5-\omega)\rho\notag\\
T^0_5&=(1+\omega)\rho u^0u_5 \notag
\end{align}
Here $u^A$ is a covariant velocity vector and $u^A u_A=1$. The
energy-momentum tensor \eqref{eq:EMTperfectfluid} corresponds to an
anisotropic perfect fluid with the isotropy parameter $w/\omega$. For
$w=\omega$ the above tensor describes isotropic perfect fluid with state
equation $p=w\rho$ where $p$ is the pressure and $\rho$ is the energy
density.

One can show that bulk Einstein equations \eqref{eq:EinsteineqnsGen} in
the generic background \eqref{eq:generalansatz} with
\eqref{eq:EMTperfectfluid} in the r.h.s. can be solved as follows
\begin{align}
\label{eq:EinsteinSolutionFluid}
a(z)&=\xi k|z|,\quad b(z)=\zeta k|z|\notag\\
\rho&=-\Lambda-6k^2 \zeta^2\notag\\
w&=-1+\frac{3\zeta^2-2\zeta\xi-\xi^2}{\Lambda/k^2+6\zeta^2}=
-1+k^2\frac{(3\zeta+\xi)(\xi-\zeta)}{\rho}\\
\omega&=-1+\frac{3\zeta(\zeta-\xi)}{\Lambda/k^2+6\zeta^2}=
-1+k^2\frac{3\zeta(\xi-\zeta)}{\rho}\notag
\end{align}
with $u_1=u_5=0$. We see that for the Lorentz invariant case
($\xi=\zeta$) only the bulk cosmological constant remains and it also
corresponds to $w=\omega=-1$. This is indeed the RS2 model but with
shifted cosmological constant $\tilde \Lambda = \Lambda+6k^2 \zeta^2$ and
appropriate matching of this constant and $k$. Therefore anisotropy of the
fluid controls the deviation from the RS2 configuration.

Israel junction conditions \eqref{eq:IsraelOur} for the brane
energy-momentum tensor \eqref{eq:energymomentumbrane} for this particular
case read
\begin{align}
-2\xi k &=p_b + \frac{2}{3}\rho_b-\frac{\sigma}{3}\notag\\
-2\zeta k &= -\frac{\rho_b}{3}-\frac{\sigma}{3}\notag
\end{align}
Combining these equations we obtain $\rho_b=6\zeta k -\sigma$ and
\begin{equation}
\label{eq:omegabrane}
\omega_b = -\frac{2}{3}-\frac{1}{3}\frac{6\xi
k-\sigma}{6\zeta k-\sigma}=-1+\frac{2k(\zeta-\xi)}{\rho_b}
\end{equation}

Let us now discuss the NECs for this configuration. One can show that for
\eqref{eq:EMTperfectfluid} in the bulk and for
\eqref{eq:energymomentumbrane} on the brane they read
%
\[
\omega\geq -1,\quad w\geq -1,\quad \omega_b
\geq -1
\]
%
We easily see that these conditions for
\eqref{eq:EinsteinSolutionFluid} and \eqref{eq:omegabrane} can be
satisfied only for the Lorentz invariant case that is for the RS2 model.

Thus we have obtained the following solution 
\[
 ds^2 = \frac{1}{k^2}\left(\frac{dt^2}{r^{2\xi}} - \frac{d\textbf{x}^2}{r^{2\zeta}}-\frac{dr^2}{r^2}\right)\,,
\]
where $r=e^{-k|z|}$. This solution exhibits the so-called Lifshitz scaling \cite{lifshitz} 
\[
 r\to \lambda r\,,\quad t\to \lambda^\xi t\,,\quad \textbf{x} \to \lambda^\zeta \textbf{x}\,,
\]
for some parameter $\lambda$.

\paragraph{The Way Out -- Positive 3d Curvature.}
Here we want to show that analogously with \cite{Cline:2001yt} it is
possible to have both NECs on the brane and in the bulk for a generic
background if the spatial 3D curvature is positive. The following
ansatz is employed
\[
ds^2 = e^{-2\xi k|z|}dt^2 -\frac{e^{-2\zeta
k|z|}}{\big(1+\frac{1}{4}\kappa^2\textbf{x}^2\big)^2}d\textbf{x}^2 -dz^2,
\]
where $\kappa$ in the the metric is the inverted 
radius of the 3-sphere. It is straightforward to plug this expression in
the Einstein equations to obtain the Einstein tensor ($z\neq 0$)
\begin{align}
G^0_0 &= 12\kappa^2 e^{2\zeta kz}-6k^2\zeta^2\notag\\
G^1_1 &= 4\kappa^2 e^{2\zeta kz}-k^2(\xi^2+2\xi\zeta+3\zeta^2)\\
G^5_5 &= 3\kappa^2 e^{2\zeta kz}-3k^2(\xi\zeta+\zeta^2)\notag
\end{align}
Let us omit the cosmological constant as it does not play a role for the
NEC. The Einstein equations read
\begin{align}
G^0_0 &= T^0_0=\rho\notag\\
G^1_1 &= T^1_1=-w\rho\\
G^5_5 &= T^5_5=-\omega\rho\notag
\end{align}
we have
\begin{equation}\label{eq:womegas}
w=-\frac{1}{3}+\frac{k^2(\xi+\zeta)^2}{6(2\kappa^2 e^{2\zeta kz}-k^2\zeta^2)},\,\quad
\omega=\frac{k^2(\xi\zeta+\zeta^2)-\kappa^2e^{2\zeta kz}}{2(2\kappa^2e^{2\zeta kz}-k^2\zeta^2)}\,.
\end{equation}
Now $\omega$ and $w$ become functions of $z$. We recall that for
$\kappa=0$ the violation of the bulk NEC for $\xi<\zeta$ took place. One
can easily see that in \eqref{eq:womegas} $w,\omega >-1$ for all $z$. One has $w \to
-\frac{1}{3}$ and $\omega\to -\frac{1}{4}$ as $z\to +\infty$ and proper choice
of $\kappa/k$ provides nonsingular behavior of $w$ and $\omega$ at small $z$.
At the same time the brane NEC holds since in this case we need merely
\eqref{eq:braneNEC} which indeed holds.

\section*{Conclusions}

In this paper we proved the strong version of the no-go theorem thereby
approving  the result of \cite{Cline:2001yt}. Our statement does not allow
generic smooth flat braneworld backgrounds with positively defined energy
density satisfying NECs in the bulk and on the brane to exist. It is
noteworthy that the statement of the theorem does not depend on finiteness
of the volume of the extra dimension which claims that the integral
\begin{equation}
\int\limits_{-\infty }^{+\infty}dz\,\sqrt{g}<\infty.\notag
\end{equation}
converges. It appears that for a generic choice of the function $a$ and
$b$ from \eqref{eq:generalansatz} corresponding to the background with
broken bulk Lorentz invariance the theorem holds. But it fails for the
Lorentz invariant setup and one can obtain any number (depending on a
particular choice of matter in the bulk) of the corresponding solutions.

Then we elaborated the one parametric Lorentz violating braneworld
solution with anisotropic perfect fluid and negative cosmological constant
in the bulk. The ani\-so\-tro\-py parameter corresponds to the parameter
controlling Lorentz invariance violation. This solution was shown to exhibit Lifshitz scaling.\footnote{Later a field theoretical solution of a background with Lifshitz scaling was found \cite{Kachru:2008yh} (see also \cite{Gordeli:2009vh}).} The well known RS2 model is a
particular case of the setup considered in this work and corresponds to
the unbroken Lorentz invariance. It was also shown that if the brane
curvature is positive then the theorem can be evaded provided that three
dimensional spatial curvature is of the same order as the Anti-de-Sitter
scale. It would mean, however, that warping effects in the extra dimension
are of the same order as curvature effects in our Universe what is
unacceptable from a phenomenological point of view.

One may expect that for time dependent ansatz with a broken Lorentz
invariance the above pattern can be dramatically improved. From
qualitative reasons one may think about fast enough changing of a scale
factor with time in order to adjust Einstein tensor to provide $T^A_B$
satisfying null energy and positiveness of energy density conditions. This
aspects will be elaborated in future works.

\acknowledgments
The authors are grateful to V. Rubakov, S. Sibiryakov, D. Levkov, D.
Gorbunov, S. Dubovsky for fruitful discussions.  M.L. would like to thank
the Service de Physique Th\'{e}orique, Universit\'{e} Libre de Bruxelles
where part of this work was done, for kind hospitality. P.K. thanks
Max-Planck-Institut f\"ur Gravitationsphysik (Albert-Einstein-Institut)
where he did part of this work for hospitality.

This work is supported in part by RFBR grant (05-02-17363-a), grant of the
President of Russian Federation (NS-7293.2006.2), grant of Dynasty
Foundation awarded by the Scientific Board of ICFPM, and  INTAS grant
(M.\,L., YSF 04-83-3015).

\end{document}